\documentclass[aps, prb, twocolumn, showpacs, floatfix]{revtex4}
\usepackage{graphicx}

\begin{document}

\title{Electronic structures of organic molecule encapsulated BN nanotubes under transverse electric field}

\author{Wei He}
\author{Zhenyu Li}
\author{Jinlong Yang}
\thanks{Corresponding author. E-mail: jlyang@ustc.edu.cn}
\author{J. G. Hou}

\affiliation{Hefei National Laboratory for Physical Sciences at
     Microscale,  University of Science and Technology of
     China, Hefei,  Anhui 230026, China}

\date{\today}

\begin{abstract}
The electronic structures of boron nitride nanotubes (BNNTs) doped
by different organic molecules under a transverse electric field
were investigated via first-principles calculations. The external
field reduces the energy gap of BNNT, thus makes the molecular bands
closer to the BNNT band edges and enhances the charge transfers
between BNNT and molecules. The effects of the electric field
direction on the band structure are negligible. The electric field
shielding effect of BNNT to the inside organic molecules is
discussed. Organic molecule doping strongly modifies the optical
property of BNNT, and the absorption edge is red-shifted under
static transverse electric field.
\end{abstract}

\pacs{75.75.+a, 73.22.-f, 72.80.Rj}

\maketitle

\section{introduction}

Nanotube is a very important actor in nanodevice applications due to
its novel properties. Compared to the more widely studied carbon
nanotube (CNT), boron nitride nanotube (BNNT) has its own
distinctive properties. Besides the excellent mechanical stiffness
and thermal conductivity comparable to CNT, \cite{review} BNNT has a
unified electronic structure regardless of its diameter, chirality,
and the number of walls of the tube. \cite{bn1, bn2, bn3} BNNT also
shows pronounced resistance to oxidation, and it is stable up to 700
$^{\circ}$C in air, while CNT survives only below 400 $^{\circ}$C.
\cite{bn5} All these properties make BNNT an attractive candidate
for nano-electronics.

Pristine BNNT is a wide-gap semiconductor. For electronics
application, it is desirable to make $p$- or $n$-type doping to
BNNT. Our previous study shows that the electronic structure of BNNT
can be modified by organic molecule encapsulation. \cite{bnhe}
Electrophilic molecule introduces acceptor states in the wide gap of
BNNT close to the valence band edge (VBE), which makes the doped
system a $p$-type semiconductor. However, with typical nucleophilic
organic molecules, instead of shallow donor states, only deep
occupied molecular states are observed. There is a significant
electron transfer from BNNT to an electrophilic molecule, while the
charge transfer between a nucleophilic molecule and BNNT is
negligible.

On the other hand, previous theoretical studies \cite{field0,
field1, field2} showed that the band gap of CNT or BNNT can be
reduced and even closed by applying a transverse electric field, due
to the giant Stark effect (GSE). \cite{field1} The electric field
can mix the nearby subband states in the valence band (VB) complex
and conduction band (CB) complex separately, leading to an electric
field dependence for the band gap. This effect is more remarkable in
BNNTs than in CNTs, due to the reduced screening of the electric
field in BNNTs. The GSE was lately confirmed experimentally using
the bias dependent scanning tunneling microscopy and scanning
tunneling spectroscopy. \cite{field3}

Since the electronic structure of BNNT can be tuned by both organic
molecule encapsulation and transverse electric field, it is thus
interesting to see what will happen if we apply both. In this paper,
by performing density functional theory (DFT) calculations, we study
the electronic structures of organic molecule encapsulated BNNTs
under a transverse electric field. The electrostatic shielding of
BNNT to the inside molecules and the effects of organic molecule
doping and electric field on the optical properties of BNNT are also
studied.

Following our previous study, \cite{bnhe} several typical
electrophilic and nucleophilic molecules are considered for BNNT
encapsulation. The two electrophilic molecules studied are
tetracyano-p-quinodimethane (TCNQ) and
tetrafluorocyano-p-quinodimethane (F4TQ). Three nucleophilic
molecules are selected: tetrakis(dimethylamino)ethylene (TDAE),
anthracene (ANTR), and tetrathia-fulvalene (TTF). The (16,0) BNNT is
chosen as a prototype in this study. We name the organic molecules
capsulated BNNT as M@BN, where M is the name of the doped organic
molecules.

\begin{figure}[!hbp]
\includegraphics[width=8cm]{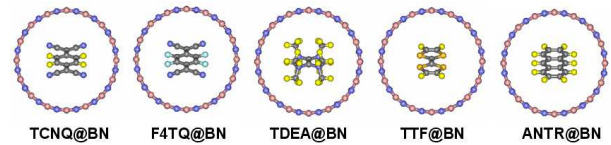}
\caption{(Color online) Optimized geometrical structures of the M@BN
systems under a 0.5 V/\AA\ electric field. B in dark-salmon, N in
blue, C in grey, H in yellow, F in aqua, and S in goldenrod. }
\label{fig:geo}
\end{figure}

The rest of the paper is organized as follows: In Sec. II, the
theoretical approach and computational details are briefly
described. In Sec. III, the calculated band structures, charge
transfers between BNNT and molecules, shielding effect under
electric field, and optical properties are discussed. Finally, we
conclude in Sec. IV.

\section{Computational Details}

To investigate the geometry structure and electronic states of the
organic molecule encapsulated BNNT, we performed first-principles
spin-polarized DFT calculations. The computational details of
geometry optimization and band structure calculation can be found in
our previous study, \cite{bnhe} and we only give a brief summary
here. We used the projector augmented-wave (PAW) method \cite{20,21}
implemented in the Vienna Ab Initio Simulation Package (VASP).
\cite{17,18} The Perdew, Burke and Ernzerhof (PBE)
exchange-correlation functional was chosen. \cite{19} We note that
more rigorous band structure calculations are required to yield
correct band gaps. \cite{Baumeier07} However, PBE functional has
been widely used in similar systems, and it should give correct
trends which are interested in this study. The energy cutoff for
plane-waves was 400 eV. In our calculation, each BNNT was separated
by 10 \AA\ of vacuum, and the minimum distance between periodically
repeated organic molecules along the BNNT tube axis was
$\sim$8.7\AA. A 4$\times$1$\times$1 $\Gamma$-centered Monhkorst-Pack
$k$-point grid was used for Brillouin zone sampling.

The way to handle electric field in VASP is adding an artificial
dipole sheet in the middle of the vacuum part in the periodic
supercell. \cite{vaspdipole1, vaspdipole2} In our study, the
direction of the electric field was chosen to be perpendicular to
the nanotube's axis (along the $y$ direction), with a magnitude of
0.20 or 0.50 V/\AA. As discussed later, the selected fields are
strong enough to make significant change of the electronic
structure, and weak enough to avoid artificial field emission.

\begin{figure}[!hbp]
\includegraphics[width=8cm]{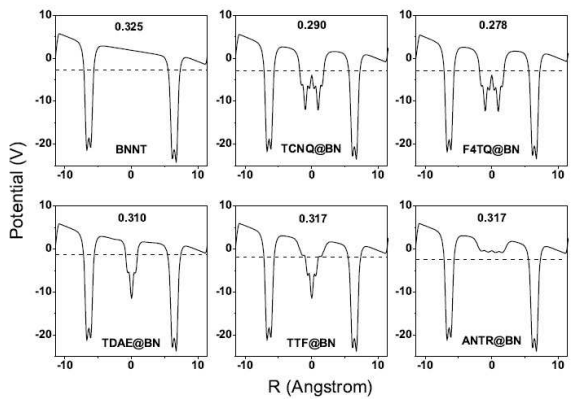}
\caption{Electrostatic potential (in volt) across the BNNT in the
M@BN systems under 0.5 V/\AA\ external electric field. The top of
the highest occupied bands are marked by dash lines. The numbers on
the top of the curves are the effective electric field (in V/\AA),
calculated by the slope of the line connecting the left
high-potential point and the right low-potential point in vacuum.}
\label{fig:potential}
\end{figure}

All linear optical properties, such as absorption coefficient and
refractive index, can be obtained as functions of dielectric
function. The real part of the dielectric function can be obtained
from its imaginary part by Kramer-Kronig transformation, and the
imaginary part $\epsilon_2$ can be calculated based on the
independent-particle approximation, \cite{Adolph0108} i.e.,
\begin{equation}
 \epsilon_2(\omega)=\frac{4\pi}{\Omega\omega^2}\sum_{i,j}\sum_\mathbf{k}
 W_\mathbf{k}|P_{ij}|^2\delta(E_{\mathbf{k}j}-E_{\mathbf{k}i}-\omega)
\end{equation}
where $\Omega$ is the unit cell volume, and $\omega$ is the photon
energy. Summations for $i$ and $j$ are performed for valence and
conduction bands, respectively. $P_{ij}$ denotes the dipole
transition matrix elements obtained from the self-consistent band
structure calculations and $W_\mathbf{k}$ is the $\mathbf{k}$ point
weighting. The same method has been applied to h-BN, and the results
are in reasonably good agreement with experiments. \cite{Guo0502}
The number of bands used for dielectric function calculations are
the sum of the electron number and the ion number. The number of
occupied bands is half of the number of electron, and the number of
unoccupied bands are more than 380 for all the systems studied in
this paper. The optical polarization in dielectric function
calculations is perpendicular to the nanotube axis. In this way, for
some parts of the tube wall, the optical polarization is nearly
perpendicular to the BN layer, while for the other parts of the tube
wall, it is roughly parallel to the BN layer.

\section{RESULTS AND DISCUSSION}

\subsection{Geometric and band structures}

As reported in our previous study, \cite{bnhe} after the organic
molecules are encapsulated into BNNT, their geometries changes
little. This suggests the weak interaction between BNNT and the
encapsulated molecules, due to the large distances between them. The
separations between neighboring molecules and between molecule and
BNNT are larger than additions of the corresponding atomic van der
Waals radius. Optimizations under the transverse electric field gave
similar structure as in the zero-field case. The optimized
geometrical structures under a 0.5 V/\AA\ electric field are shown
in Fig. \ref{fig:geo}. The relaxed geometries of the M@BN systems
before and after applying electric field change little. The largest
change of bond length, comparing to the geometry under zero field,
is smaller than 0.01 \AA. Consistently, the direction of the
transverse field does not affect the geometric and electronic
structures. Our test calculation gave almost identical band
structures for F4TQ@BN under 0.5 V/\AA\ transverse electric field
along two different directions. For simplicity, in the rest part of
the paper, all electronic structure calculations were done based on
the geometrical structures obtained without electric field.

Before calculating the electronic structure of M@BN with transverse
electric field applied, two important issues are needed to be
discussed. One is that the vacuum region should be broad enough to
avoid the overlap between the charge density and the artificial
dipole sheet used to generate electric field. The width of the
vacuum layer between two neighboring BNNTs in our calculation is 10
\AA, and the work function of (16,0) BNNT is 6.36 eV. The charge
density of BNNT roughly decay as $\exp(-\sqrt{\phi}r)$, where $\phi$
is the work function in eV and $r$ is the distance from the BNNT in
\AA. \cite{vaspdipole3} The charge density thus decays one magnitude
every 0.91 \AA. The dipole sheet is 5 \AA\ far from the BNNT, and
there is thus only negligible charge density there. The other issue
is that the vacuum region should not be too broad so as to attract
electrons to vacuum (field emission). In Fig. \ref{fig:potential},
we plot the electrostatic potential of the M@BN system under a 0.5
V/\AA\ electric field, the strongest electric field used in our
calculations. We can see that the lowest potential in vacuum region
is still higher than the edge of the highest occupied band (HOBand),
which confirms that the artificial field emission has not been
introduced in this study.

\begin{figure}[!hbp]
\includegraphics[width=8cm]{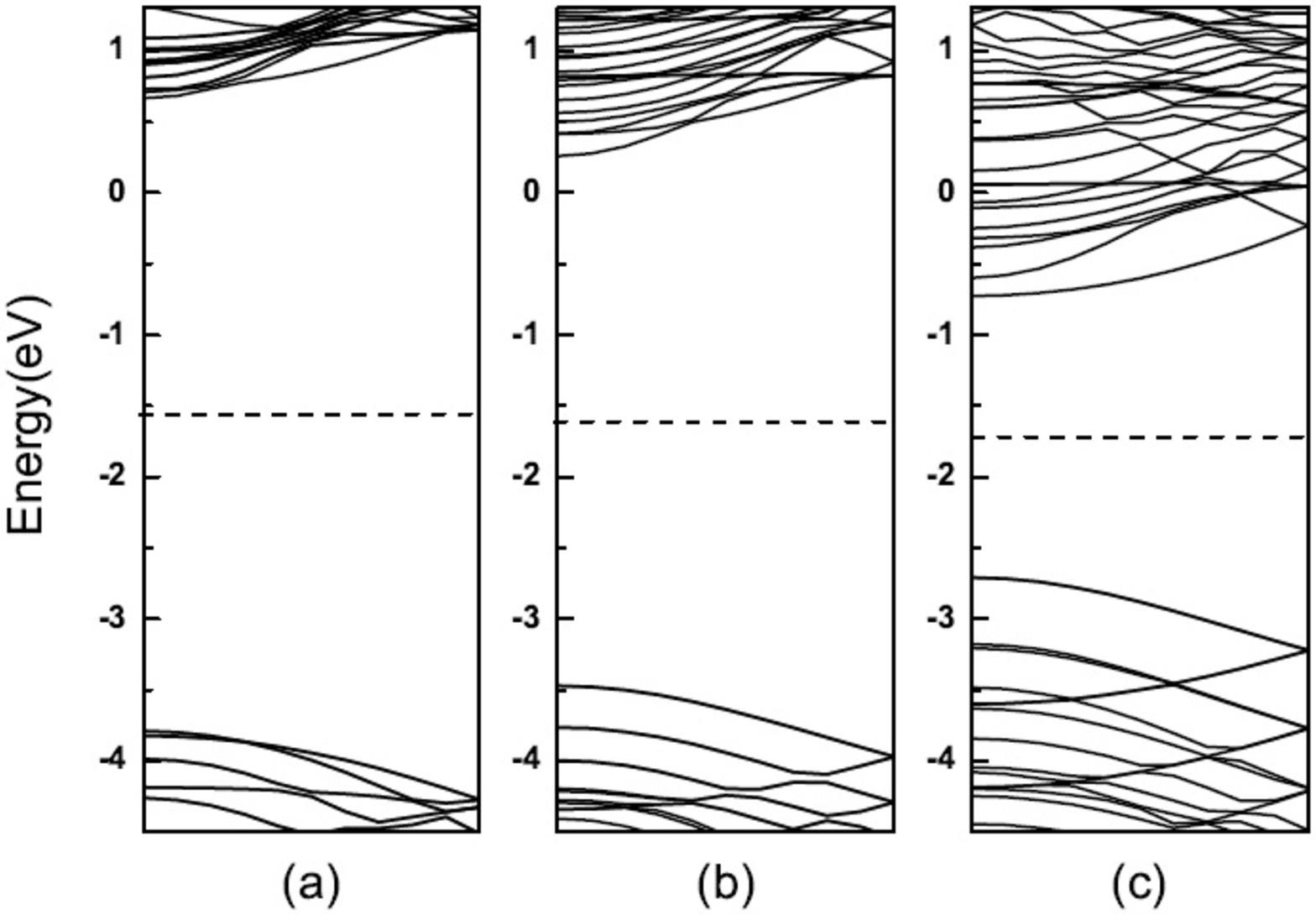}
\includegraphics[width=8cm]{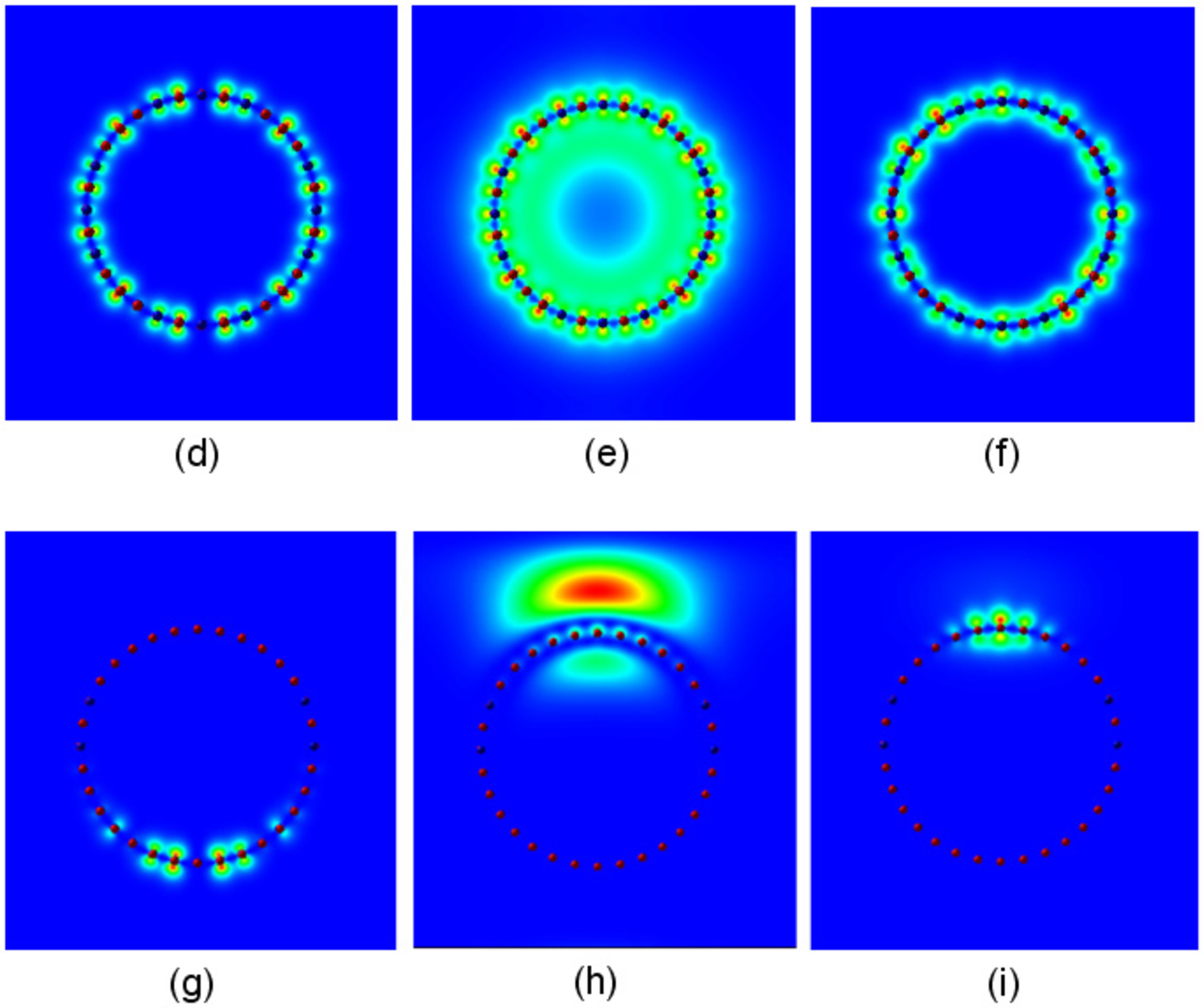}
\caption{Band structures (from $\Gamma$ to X) of the pristine BNNT
under (a) 0.0, (b) 0.2, and (c) 0.5 V/\AA\ electric fields. Fermi
levels are marked by dashed lines. Density distribution for HOBand,
LUBand, and LUBand+1 in pristine BNNT under (d-f) 0.0 and (g-i) 0.5
V/\AA\ electric fields. The electric field direction is from up to
down.} \label{fig:bandbn}
\end{figure}

We first calculate the band structure of the pristine BNNT under the
0.2 and 0.5 V/\AA\ electric fields, and the results are shown in
Fig. \ref{fig:bandbn}. The band gap decreases from 4.45 to 3.73 and
1.98 eV, when the field increases from 0 to 0.2 and 0.5 V/\AA,
respectively. Our result is consistent with the previous theoretical
study on similar systems. \cite{field1} Such band gap decrease is a
result of the extensive mixing among the subband states within the
valence and conduction band complex under electric filed. The band
mixing is also reflected by the band-resolved charge density
distributions, as shown in Fig. \ref{fig:bandbn}, where the density
of the HOBand, the lowest unoccupied band (LUBand), and LUBand+1
under 0.0 and 0.5 V/\AA\ field are plotted. The orbital density move
along the electric field for occupied states, and against the field
direction for unoccupied states. From the density distributions, the
LUBand state can be identified as a nearly free electron (NFE)
state. \cite{field1}

\begin{figure}[!hbp]
\includegraphics[width=8cm]{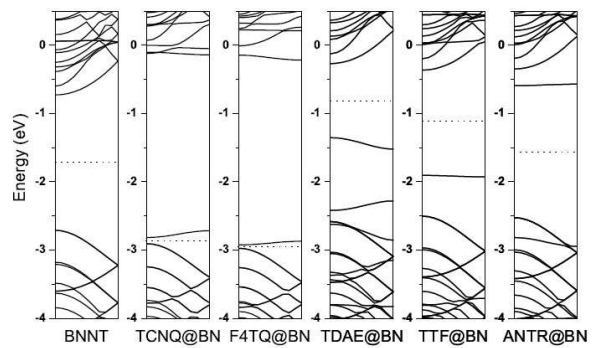}
\caption{Band structures (from $\Gamma$ to X) of the pristine BNNT
and the M@BN systems under 0.5 V/\AA\ electric field. Fermi levels
are marked by dashed lines. } \label{fig:band0.5}
\end{figure}

The band structures of the M@BN systems under 0.2 and 0.5 V/\AA\
electric field are then calculated. Similar to results under zero
electric field, \cite{bnhe} the bands are either mainly from
molecules or mainly from BNNT. The molecular bands are much flatter
than the BNNT bands. Since there is no qualitative difference
between the band structure of the M@BN systems under 0.2 and 0.5
V/\AA\ electric fields, we will focus on the 0.5 V/\AA\ case (Fig.
\ref{fig:band0.5}). For the two electrophilic molecules (TCNQ and
F4TQ), the lowest unoccupied molecular bands become very close to
the BNNT VBE, which means good $p$-type doping. It also suggests a
considerable charge transfer between the molecules and the BNNT. We
define the gap between the lowest unoccupied molecular band and the
BNNT VBE as $E_p$, and the gap between the highest occupied
molecular band and the BNNT conduction band edge (CBE) as $E_n$.
Their values are listed in Table \ref{tbl:eg}. For the nucleophilic
organic molecules TDAE and TTF, $E_n$ drops from 2.30 and 2.85 eV at
zero field \cite{bnhe} to 1.08 and 1.54 eV, respectively. However,
it is still too large to form good $n$-type doping. For ANTR@BN, the
highest occupied molecular band is pushed into the BNNT valence band
manifest, and there is only an unoccupied molecular band close to
the BNNT CBE. Therefore, neither $n$-type or $p$-type doping is
introduced to BNNT by ANTR encapsulation.

\begin{table}[b]
\caption{Energy gaps $E_p$, $E_n$, and $E_g^{BN}$ of the pristine
BNNT and the M@BN systems under electric field $\mathcal{E}$. The
energy gaps are in eV, and the electric field $\mathcal{E}$ is in
V/\AA. } \label{tbl:eg}
\begin{tabular}{cccccccc}
 \hline\hline
 Gap & $\mathcal{E}$ & BN & TCNQ & F4TQ & TDAE & TTF & ANTR   \\
 \hline
$E_p$      & 0.0 &  --- & 0.15 & 0.09 & 4.37 & 3.55 & 2.96 \\
           & 0.2 &  --- & 0.17 & 0.11 &  --- & 3.25 & 2.65 \\
           & 0.5 &  --- & 0.09 & 0.05 &  --- &  --- & 1.93 \\
$E_n$      & 0.0 &  --- &  --- &  --- & 2.30 & 2.85 & 3.71 \\
           & 0.2 &  --- &  --- &  --- & 1.86 & 2.49 & 3.40 \\
           & 0.5 &  --- &  --- &  --- & 1.08 & 1.54 &  --- \\
$E_g^{BN}$ & 0.0 & 4.45 & 4.45 & 4.48 & 4.45 & 4.46 & 4.43 \\
           & 0.2 & 3.73 & 4.00 & 4.18 & 3.63 & 3.79 & 3.81 \\
           & 0.5 & 1.98 & 2.79 & 2.91 & 2.31 & 2.14 & 2.18 \\
 \hline\hline
\end{tabular}
\end{table}

To consider the molecular effect on the intrinsic electronic
structure of BNNT under electric field, we define the energy gap
between the CBE and VBE of BNNT as $E_g^{BN}$. As shown in Table
\ref{tbl:eg}, without external field, it is not very sensitive to
the type of molecules. However, $E_g^{BN}$ for TCNQ@BN and F4TQ@BN
under electric field is much larger than other systems under the
same field. This can be easily understood by their electronic
structures. Under strong external field, the energy gaps of TCNQ@BN
and F4TQ@BN become very small, which means enhanced metallicity and
stronger electrostatic screening. With stronger screening, the
reduction of electrostatic potential along the direction of electric
field is slower (see Fig. \ref{fig:potential}), therefore,
$E_g^{BN}$ reduction compared to pristine BNNT and other M@BN
systems is smaller. Similar behavior has been observed in our study
on defective BNNTs under an electric field. \cite{bnhu}

\subsection{Charge transfer}

An important property of the molecule encapsulated system is the
charge transfer between the molecules and BNNT. We calculate the
charge transfer using the method described in our previous work,
\cite{bnhe} originally proposed by Lu et al..\cite{charge1} Briefly
speaking, the boundary between the organic molecule and the BNNT is
determined by the maximum/minimum position of the
cylinder-integrated differential electron density curve, and the
corresponding maximum/minimum value is the charge transfer value.
The differential electron density is defined as the difference
between electron densities of the M@BN system under electric field
and the sum of charge densities of BNNT and molecules without
electric field. The range of the radial position $R_b$ of boundary
between molecule and BNNT can be estimated by the density
distribution of the BNNT bands. For electrophilic molecules, $R_b$
should be inside the region where the BNNT HOBand has a high density
($R_b<6$ \AA), while for nucleophilic molecules, $R_b$ should be
inside region where the BNNT LUBand has a high density ($R_b<3.5$
\AA). \cite{bnhe}

\begin{figure}[!hbp]
\includegraphics[width=8cm]{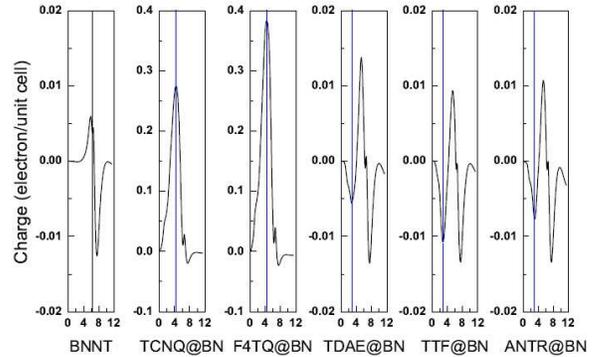}
\caption{(Color online) The cylinder-integrated differential
electron density curves for the the pristine BNNT and the M@BN
systems under a 0.5 V/A electric field. The blue vertical lines
indicate the radial boundary between molecules and the BNNT, in the
viewpoint of charge transfer. The black vertical line in the first
panel marks the radius of the BNNT.} \label{fig:E0.5charge}
\end{figure}

For pristine BNNT, the differential electron density is just the
density change before and after applying electric field. As shown in
Fig. \ref{fig:E0.5charge}, the electron density increases inside the
nanotube wall, while it decrease outside the wall. Of course, the
absolute value of this change is very small ($\sim$0.01 electron).
For the electrophilic organic molecules TCNQ and F4TQ, under 0.5
V/\AA\ electric field, there are 0.28 and 0.38 electrons per
molecule transfer from BNNT, about twice as the value without field
(0.10 and 0.24 electron). \cite{bnhe} This is due to the closer
molecular state in energy to VBE. The boundary radius is 4.3 \AA,
which is basically the midpoint between the molecule and the BNNT as
in the no field case. For the nucleophilic organic molecules TDAE,
TTF and ANTR, the charge transfers form molecules to BN also
increase, which is 0.006, 0.011, and 0.008 electron per molecule,
respectively (0.004, 0.008, and 0.005 electrons without external
field). Such small electron transfers also indicate that the doped
system is not a good $n$-type semiconductor.

\subsection{Shielding effect}

\begin{figure}[!hbp]
\includegraphics[width=8cm]{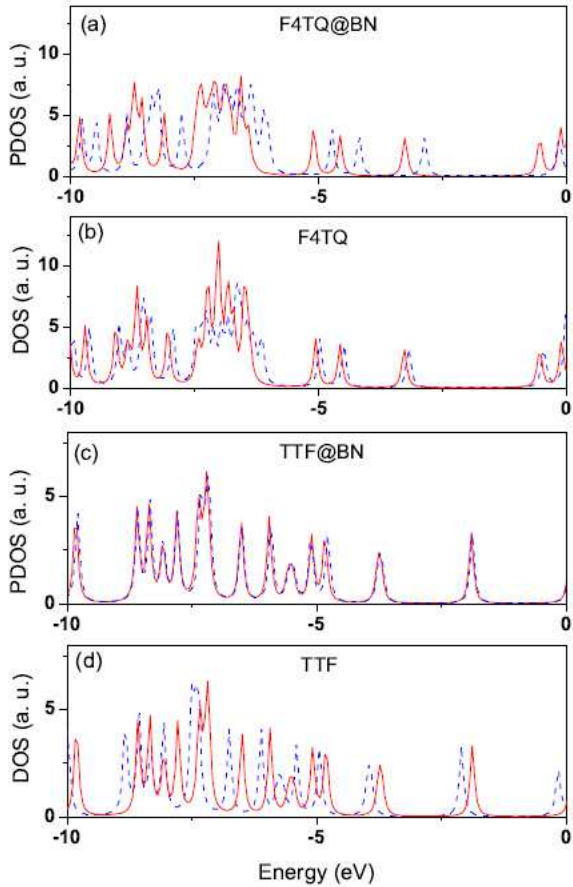}
\caption{(Color online) Molecular PDOS of (a) F4TQ@BN and (c) TTF@BN
under 0.0 (solid) and 0.5 (dashed) V/\AA\ electric field. DOS of
individual (b) F4TQ and (d) TTF molecules under 0.0 (solid) and 0.5
(dashed) V/\AA\ electric field.} \label{fig:dos}
\end{figure}

It is interesting to study the effects of BNNT to the geometry and
electronic structure of the inside molecules under an electric
field. In Fig. \ref{fig:dos}, we compare the molecular partial
densities of states (PDOS) with and without external field for two
representative molecules: an electrophilic one (F4TQ) and a
nucleophilic one (TTF). The calculated PDOS shows different
behaviors for these two types of molecules. For the electrophilic
organic molecules F4TQ (Fig. \ref{fig:dos}a), there is an nearly
rigid upshift of the molecular PDOS after applying the electric
field. This is due to the electron transfer from BNNT to the
molecules. For the nucleophilic organic molecules TTF (Fig.
\ref{fig:dos}c), the molecular PDOS almost does not change after
applying the electric field, which is consistent with the little
charge transfers between BNNT and the molecules.

We also check the shielding effect by applying electric field to
individual molecules without a BNNT encapsulation. First, we relax
the geometry of the individual molecules under the 0.5 V/\AA\
electric field. Contrasting to the encapsulated case, we found a
significant geometry relaxation after applying the electric field.
Then, we fix the molecular geometry to the one as inside BNNT, and
calculate the electronic structure under zero and finite electric
field. The effect of electric field to the molecular electronic
structures for individual molecules is different from that for
molecules encapsulated by BNNT (see Fig. \ref{fig:dos}). For F4TQ
molecule, the DOS energy shift caused by electric field is smaller
than that for PDOS of F4TQ@BN. However, we can see a notable change
of the overall DOS shape after applying the electric field. For TTF,
electric field shifts the DOS of individual molecules, but it does
not change the molecular PDOS. In summary, our results indicates,
although not good as CNT, the BNNT still provides a significant
electrostatic shielding for the inside molecules: the geometry of
the inside molecules and the electronic structures of nucleophilic
molecules are almost not affected by an external field under the
protection of BNNT.

\subsection{Optical properties}

\begin{figure}[!hbp]
\includegraphics[width=8cm]{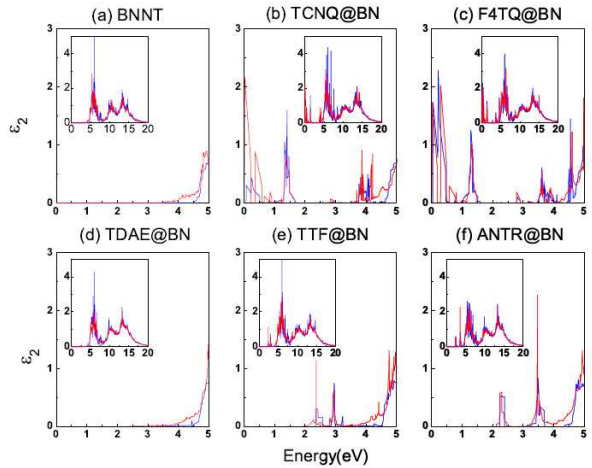}
\caption{(Color online) The imaginary part ($\epsilon_2$) of the
theoretical dielectric function of the BNNT and the M@BN systems,
with (red) and without (blue) the external static electric field.
Inset: $\epsilon_2$ plotted in a larger energy scale. }
\label{fig:optics}
\end{figure}

The calculated imaginary part of the dielectric function of the
pristine BNNT without external electric field is shown in the inset
of Fig. \ref{fig:optics}a. A much denser $k$-point sampling is
required to obtain smooth curves. \cite{Guo0502} However, the main
features of $\epsilon_2$ are already presented in our results. And
our calculated $\epsilon_2$ of prinstine BNNT agrees well with the
previous theoretical result on the zigzag BNNTs. \cite{Guo0502}
Mainly, there are three peaks at about 5.6, 10.6, and 13.5 eV,
respectively. The strong peak around 5.6 eV corresponds to the
$\pi\rightarrow\pi^*$ interband transitions. The two other peaks
higher in energy and weaker in intensity are associated with
interband transitions also involving the $\sigma$ bands. We note
that the peak around 10.6 eV does not appear in our test calculation
with the optical polarization along the tube axis. By optical
selection rule analysis, \cite{Guo0502} the 10.6 eV peak should
correspond to $\pi\rightarrow\sigma^*$ and $\sigma\rightarrow\pi^*$
transitions, and the 13.5 peak mainly contributed by
$\sigma\rightarrow\sigma^*$ transitions. In Fig. \ref{fig:optics}a,
we also plot $\epsilon_2$ of the pristine BNNT under transverse
electric field. Similar to a previous study on pristine BNNT,
\cite{field5} the whole imaginary dielectric function is not
strongly affected by the static field, but the absorption edge is
red-shifted.

With molecule intercalation, all the three peaks are still clearly
shown in the M@BN system. However, new features appear in the
low-energy region from 0 to 5 eV. These peaks below the BNNT
absorption edge ($\sim$4.5 eV) are clearly from transitions between
molecular bands and the BNNT bands or between molecular bands
themselves. For the two electrophilic molecules, there are peaks
close to the zero energy. This is consistent with the very narrow
band gap calculated for these two systems. There is no significant
peaks below 4.5 eV for TDAE@BN, and the new peaks for TTF@BN and
ANTR@BN are close to the BNNT absorption edge. Our results indicate
that the optical property of BNNT can be strongly modified by
different organic molecule encapsulation.

Next, we study the case where both molecule intercalation and
electric field exist. The theoretical imaginary part of the
dielectric functions of the molecule doped BNNT under a 0.5 V/\AA\
transverse static electric field are also presented in Fig.
\ref{fig:optics}. The three peaks from BNNT are still not strongly
affected by the static field. However, if we look into the details
below 5 eV, we can see three different behaviors. For transitions
between valence band and conductance bands of BNNT, the electric
field GSE broadens the band manifests, thus broadens the absorption
peaks too. This broadening effect red-shifts the absorption edge of
BNNT as shown in Fig. \ref{fig:optics}d. For transitions between
molecular bands, due to electrostatic shielding of BNNT, the
electric field does not strongly affect the peaks, as shown in Fig.
\ref{fig:optics}e and \ref{fig:optics}f. For transitions between
BNNT bands and molecular bands, the electric field affects both the
amplitude of the peaks and the peak positions, as shown in Fig.
\ref{fig:optics}b and \ref{fig:optics}c.

\section{conclusion}

In summary, we calculate the electronic structure and optical
properties of BNNT under transverse electric field with different
molecule intercalations. When applying transverse electric field,
the band gap of BNNT decrease, while the band structures of organic
molecules change little, or only with a rigid shift. After applying
electric field, there is more charge transfer between molecules and
nanotube in the $p$-type semiconductors TCNQ@BN and F4TQ@BN. BNNT
provides relatively good electrostatic shielding for the geometry
relaxation of the inside organic molecular chain, and for the
electronic structure of nucleophilic molecules. Organic molecule
doping strongly modifies the optical properties of the composite
M@BN systems. The absorption edge is slightly red-shifted under
static electric field.

\begin{acknowledgements}
The postprocessing routine downloaded from the homepage of Prof.
Furthmuller has been used in the optical property calculations. This
work is partially supported by the National Natural Science
Foundation of China (50721091, 20533030, 50731160010), by National
Key Basic Research Program under Grant No.2006CB922004, by the
USTC-HP HPC project, and by the SCCAS and Shanghai Supercomputer
Center.
\end{acknowledgements}


\begin{thebibliography}{99}

\bibitem{review} D. Golberg, Y. Bando, C. C. Tang, and C. Y. Zhi, Adv. Mater. {\bf 19}, 2413(2007).

\bibitem{bn1} A. Rubio, J. L. Corkill, and M. L. Cohen, Phys. Rev. B  {\bf 49}, 5081 (1994).

\bibitem{bn2} N. G. Chopra, R. J. Luyken, K. Cherrey, V. H. Crespi, M. L. Cohen, S. G. Louie, and A. Zettl, Science {\bf 269}, 966 (1995).

\bibitem{bn3} X. Blase, A. Rubio, S. G. Louie, and M. L. Cohen, Europhy. Lett. {\bf 28}, 335 (1994).


\bibitem{bn5} Y. Chen, J. Zou, S. J. Campbell, and G. L. Caer, Appl. Phys. Lett. {\bf 84}, 2430 (2004).

\bibitem{bnhe} W. He, Z. Li, J. Yang, and J. G. Hou, J. Chem. Phys., 128, 164701 (2008).

\bibitem{field0} J. O'Keeffe, C. Wei, and K. Cho, Appl. Phys. Lett. {\bf 80}, 676 (2002).

\bibitem{field1} K. H. Khoo, M. S. C. Mazzoni, and S. G. Louie, Phys, Rev. B {\bf 69}, 201401 (2004)

\bibitem{field2} C. W. Chen, M. H. Lee, adn S. J. Clark, Nanotechnology {\bf 15}, 1837 (2004)

\bibitem{field3} M. Ishigami, J. D. Sau, S. Aloni, M. L. Cohen, and A. Zettl, Phys, Rev. Lett. {\bf 94}, 056804 (2005)

\bibitem{20} G. Kresse, D. Joubert, Phys. Rev. B {\bf 59}, 1758 (1999).

\bibitem{21} P. E. Blochl, Phys. Rev. {\bf 50}, 17953 (1994).

\bibitem{17} G. Kresse, J. Furthm¨¹ller, Phys. Rev. B {\bf 64}, 11169 (1996).

\bibitem{18} G. Kresse, J. Furthmuller, Comput. Mater. Sci. {\bf 6}, 15 (1996).

\bibitem{19} J. P. Perdew, K. Burke, and M. Ernzerhof, Phys. Rev. Lett. {\bf 77}, 3865 (1996).

\bibitem{Baumeier07} B. Baumeier, P. Kruger, and J. Pollmann, Phys.
Rev. B {\bf 76}, 085407 (2007).

\bibitem{vaspdipole1} J. Neugebauer and M. Scheffler, Phys. Rev. B 46, 16067 (1992).

\bibitem{vaspdipole2} G. Kresse and J. Furthmuller, VASP Guide.

\bibitem{Adolph0108} B. Adolph, J. Furthmuller, and F. Bechsted,
Phys. Rev. B {\bf 63}, 125108 (2001)

\bibitem{Guo0502} G. Y. Guo and J. C. Lin, Phys. Rev. B {\bf 71},
165402 (2005)

\bibitem{vaspdipole3}P. J. Feibelman, Phys. Rev. B 64, 125403(2001).

\bibitem{bnhu} S. Hu, Z. Li, X. C. Zeng, and J. Yang, J. Phys. Chem.
C, ASAP article (2008)

\bibitem{charge1} X. Lu, M. Grobis, K. H. Khoo, S. G. Louie, and M. F. Crommie, Phys. Rev. B 70, 115418 (2004).




\bibitem{field5} C. W. Chen, M. H. Lee, and Y. T. Lin, Appl. Phys. Lett. {\bf 89}, 223105 (2006).





\end{thebibliography}
\end{document}